\documentclass[preprint,showpacs,preprintnumbers,amsmath,amssymb,superscriptaddress]{revtex4}
     
\usepackage{graphicx}
\usepackage{dcolumn}
\usepackage{bm}

\begin{document} 

\preprint{Manuscript: Mechanically encoded single photon sources}

\title{
Mechanically induced pseudo-magnetic fields in the excitonic fine structures of droplet epitaxial quantum dots}

\author{Shun-Jen Cheng}
\affiliation{Department of Electrophysics, National Chiao Tung
  University, Hsinchu 30050, Taiwan, Republic of China.
}
\author{Yu-Huai Liao}
\affiliation{Department of Electrophysics, National Chiao Tung
  University, Hsinchu 30050, Taiwan, Republic of China.
}
\author{Pei-Yi Lin}
\affiliation{Department of Electrophysics, National Chiao Tung
  University, Hsinchu 30050, Taiwan, Republic of China.
}


\date{\today}

\begin{abstract}
We present numerical investigations based on the Luttinger-Kohn four-band $k \cdot p$ theory and, accordingly, establish a quantitatively valid model of the excitonic fine structures of droplet epitaxial GaAs/AlGaAs quantum dots under uni-axial stress control.
In the formalisms, stressing a photo-excited quantum dot is equivalent creating a pseudo-magnetic field that is directly coupled to the pseudo-spin of the exciton doublet and tunable to tailor the polarized fine structure of exciton.    
The latter feature is associated with the valence-band-mixing of exciton that is especially sensitive to external stress in inherently unstrained droplet epitaxial GaAs/AlGaAs quantum dots and allows us to {\it mechanically} design and prepare any desired exciton states of QD photon sources prior to the photon generation.  
\end{abstract}

\maketitle

\section{Introduction}

Excitonic fine structures (FS's) of semiconductor quantum dots (QDs) have been realized as an essential feature of advanced photonic applications, such as entangled photon pair emitters \cite{Benson,Bennett2010,Mohan,Trotta,Kuroda} and exciton-qubit gates. \cite{Benny, Kodriano} The realization of entangled photon pair emitters based on QDs has been for a long time a challenging task because it needs to retain the degeneracy of exciton doublet states, which is however likely lift by any slight symmetry breakings of QD structure. \cite{Singh,Bryant,Seguin,Gong,Ramirez}
By contrast to the application of entangled photon pair generation, an efficient operation of exciton qubit gate yet needs an energy level anti-crossing in the FS that is coherently tunable so that any desired superposition states can be deterministically prepared and controlled.\cite{Pooley}

Technologically, Trotta {\it et al.} have recently demonstrated an efficient way to retain, universally, the degeneracies of exciton doublet states of asymmetric QDs by electrical and mechanical means.\cite{Trotta} The success in the exploitation of mechanical stress control paves an inspiring way to extend the usefulness of QD photon emitters with the potential integrations with micro electro-mechanical systems (MEMS)\cite{Baek} and nano-acoustics.\cite{Gust,Kataoka} In the realization of those scaled-up hybrid quantum systems, a crucial issue is if whether and to what extent the quantum nature of a QD device can be affected by the applied mechanical stresses. 

This work presents numerical investigations based on the Luttinger-Kohn four-band $k \cdot p$ theory and, accordingly, establishes a quantitatively valid model of excitonic fine structures of droplet epitaxial (DE) GaAs/AlGaAs QDs under uni-axial stress control. As a main feature elucidated by our studies, imposing an external stress onto a QD is shown not only to alter the magnitude of fine-structure splitting (FSS) but also rebuild the coherent superposition of exciton states significantly.\cite{Singh,Gong} The latter feature is associated with the valence-band-mixing (VBM) of exciton that is especially sensitive to external stress in inherently unstrained DE-QDs and allows us to design and prepare, {\it mechanically}, desired exciton states of a QD photon source prior to phonon emission.  In the model, we formulate an uniaxial stress applied on a photo-excited quantum dot as a pseudo-magnetic field that is directly coupled to the pseudo-spin of the exciton doublet and tunable to change the level splitting and the coherent superposition of the exciton states. The concept of such a stress-induced pseudo-magnetic field has been explored extensively in the field of two-dimensional monolayer materials, e.g. graphenes, very recently \cite{Guinea, Juan} and is demonstrated here to be crucial as well in quasi-zero-dimensional systems.  Furthermore, photon pairs emitted from stress-controlled vanishing fine structure splitting (FSS) are predicted to be always {\it non-maximally} entangled (also referred to as hyper-entanglement), an useful feature for loophole-free tests of Bell inequality.\cite{White, Eberhard} Those revealed features that are beyond the most existing schemes simply based on pure heavy-hole-exciton are well captured by our improved model with the thorough consideration of the VBM nature of exciton.

\section{Theoretical framework}
\label{theoryframe}

We begin with the Hamiltonian for an interacting exciton in a QD that is expressed in the language of second quantization as, 
$ H_X  = \sum_{i_e} E_{i_e}^e c_{i_e}^+c_{i_e}
+ \sum_{i_h} E_{i_h}^h h_{i_h}^+h_{i_h} 
-  \sum_{i_e,j_h,k_h,l_e}V_{i_e,j_h,k_h,l_e}^{eh}c_{i_e}^+h_{j_h}^+h_{k_h}c_{l_e} + \sum_{i_e,j_h,k_h,l_e}V_{i_e,j_h,k_h,l_e}^{eh,xc}c_{i_e}^+h_{j_h}^+h_{k_h}c_{l_e} $, \cite{RamirezPRL} 
where $i_{e}$ ($i_{h}$) represents a composite index composed of the labels of orbital and spin of a single-electron (single-hole) state, $c_{i_e}^+$ and $c_{i_e}$ ($h_{i_h}^+$ and $h_{i_h}$) are the particle creation and annihilation operators, 
$V_{i_e,j_h,k_h,l_e}^{eh}\equiv \int \int d^3{r_{e}} d^3{r_{h}} \psi_{i_e}^{e\ast}(\vec{r_{1}}) \psi_{j_h}^{h\ast}(\vec{r_{2}}) \frac{e^{2}}{4 \pi \epsilon_0 \epsilon_b |\vec{r}_{12}|}  \psi_{k_h}^{h}(\vec{r_{2}}) \psi_{l_e}^{e}(\vec{r_{1}})$
($V_{i_e,j_h,k_h,l_e}^{eh,xc}\equiv \int \int d^3{r_{1}} d^3{r_{2}} \psi_{i_e}^{e\ast}(\vec{r_{2}}) \psi_{j_h}^{h}(\vec{r_{2}}) \frac{e^{2}}{4 \pi \epsilon_0 \epsilon_b |\vec{r}_{12}|}  \psi_{k_h}^{h \ast}(\vec{r_{1}}) \psi_{l_e}^{e}(\vec{r_{1}}) $) are the matrix elements of conventional electron-hole ({\it e-h}) Coulomb interactions ({\it e-h exchange} interactions), $\vec{r}_i$ denotes the coordinate position of particle, $\vec{r}_{12}\equiv \vec{r}_1- \vec{r}_2$, $\epsilon_0$ is vacuum permittivity, $\epsilon_b=\epsilon_b(|\vec{r}_{12}|)$ is the dielectric function of material that is generally dependent on the inter-particle distance, $E_{i_e}^e$ and $E_{i_h}^h$ ($\psi_{i_e}^{e}$  and $\psi_{i_h}^{h}$) are the eigen energies (wave functions) of a single electron and single hole in the QD, respectively.  For the wide-band-gap GaAs/AlGaAs QDs studied in this work, we study the single-electron (single-hole) spectra, $\{E_{i_e}^e\}$ ($\{ E_{i_h}^h\}$), of a QD in the framework of the single band model (four-band $k\cdot p$ model), and the single-electron (-hole) wave functions are written as $\psi_{i_e}^{e}(\vec{r}_e)=g_{i_e}^e(\vec{r}_e) u_{s_z}^e(\vec{r}_e)$ ($\psi_{i_h}^{h}(\vec{r}_h)=\sum_{j_z=\pm\frac{1}{2},\pm\frac{3}{2}}g_{i_h,j_z}^h(\vec{r}_h) u_{j_z}^h(\vec{r}_h)$), a product of slowly varying envelope functions $g_{i_e}^{e}$ ($\{ g_{i_h,j_z}^h\}$) and microscopic Bloch functions, $u_{s_z}^{e}$ ($u_{j_z}^h$), of spin $s_z=\pm\frac{1}{2}$ (of angular momenta $j_z=\pm \frac{1}{2},\pm \frac{3}{2}$) for a conduction electron (valence hole). 

\begin{figure}[t]
\includegraphics[width=12.0cm]{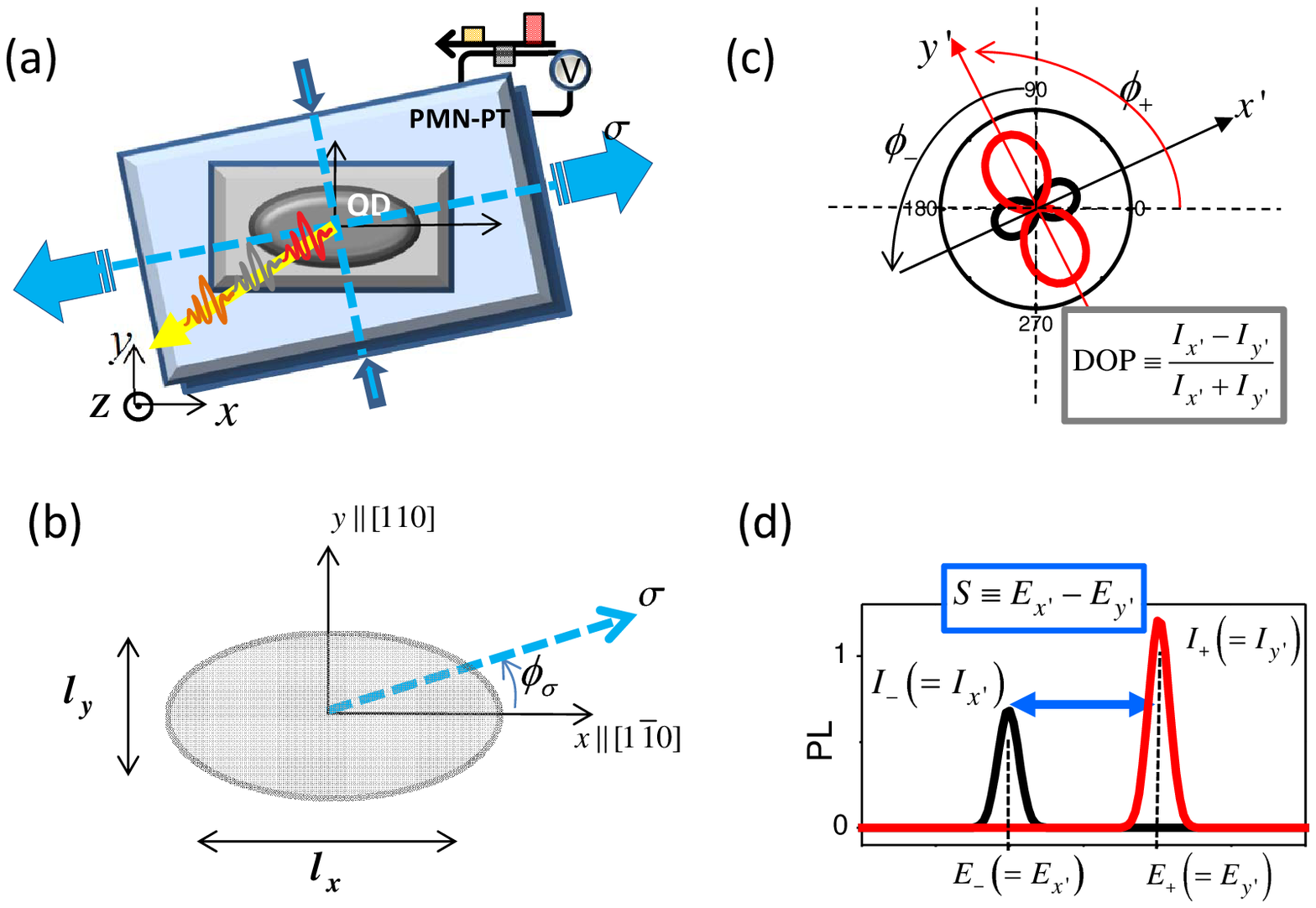}
\caption{(a) Schematics of a quantum dot (QD) photon source mounted on a piezoelectric actuator (PMN-PT) under a controlled uniaxial stress $\sigma$ along the direction with a angle $\phi_{\sigma}$ to the elongation axis ($x$-axis) of the QD. Throughout this work, we consider $0^\circ <\phi_{\sigma}< 45^\circ$. (b) Polarized fine structures of exciton of a stressed QD without and with valence band mixing (VBM). (c)  Polar plot of optical polarization and (d) polarized emission spectrum of a VBM exciton in a uniaxially stressed QD.}
\label{fig1}
\end{figure}

In the theoretical framework, the envelope wave function, $g_{i_e}^e$, of a single electron in a stressed QD satisfies the Schr\"odinger equations, $H_{e}g_{i_e}^e=E_{i_e}^e g_{i_e}^e$, where 
\begin{equation}
H_e=\frac{\hbar^2 (k_x^2 + k_y^2 + k_z^2)}{2m_e^\ast}+V_{QD}^e(\vec{r}_e)+a_c(\epsilon_{xx}+\epsilon_{yy}+\epsilon_{zz})
\end{equation}
is the single-electron Hamiltonian in the single-band effective mass approximation, $\epsilon_{\alpha\beta}$ are the tensor elements of strain ($\alpha,\beta=x,y,z$), $k_{\alpha} = -i\frac{\partial }{\partial \alpha}$ is the operator of the $\alpha$-component of wave vector, $V_{QD}^e(\vec{r}_e)$ is the position-dependent confining potential for an electron in the dot, $m_e^\ast =0.067m_0$ is the effective mass of electron, $m_0$ is the free electron mass, and  $a_c=-8.013$eV for GaAs.\cite{Chuang}

Within the four-band Luttinger-Kohn $k\cdot p$ model, the Hamiltonian for a single hole in the same stressed QD is formulated as a $4\times 4$ matrix, $H_{h}=H_k^h+H_{\epsilon}^h+V_{QD}^h I_{4\times 4}$, that is composed of the kinetic energy-, strain- and potential parts, respectively. The single hole spectrum of a QD is calculated by solving $H_h |\psi_{i_h}^{h}\rangle = E_{i_h}^h |\psi_{i_h}^{h}\rangle $.  In the basis ordered by $\{ |u_{\frac{3}{2}}\rangle, |u_{\frac{1}{2}} \rangle,|u_{-\frac{1}{2}}\rangle,|u_{-\frac{3}{2}}\rangle   \}$, the single-hole wave functions are expressed as 4-vectors, $|\psi_{i_h}^{h}\rangle =(g_{i_h,\frac{3}{2}}^h (\vec{r}_h), g_{i_h,\frac{1}{2}}^h (\vec{r}_h), g_{i_h,-\frac{1}{2}}^h (\vec{r}_h) g_{i_h,-\frac{3}{2}}^h (\vec{r}_h))$, and the kinetic energy part of the Hamiltonian is expressed as
\begin{equation}
\label{hammat}
 H_k^h = 
\left(
\begin{array}{cccc}
P_{k}+Q_{k} & -S_k & R_k & 0 \\
-S_k^+ & P_{k}-Q_{k} & 0 & R_k \\
R_k^+ & 0 & P_{k}-Q_{k} & S_k \\
0 & R_k^+  & S_k^+ & P_{k}+Q_{k}
\end{array}
\right) \, ,
\end{equation}
where $P_k = \frac{\hbar^2\gamma_1}{2m_0}\left(k_x^2+k_y^2+k_z^2\right)$, $Q_k = \frac{\hbar^2\gamma_2}{2m_0}\left(k_x^2+k_y^2-2k_z^2\right)$, 
$R_k = \frac{\hbar^2}{2m_0}\left[-\sqrt{3}\gamma_3\left(k_x^2-k_y^2\right)+i2\sqrt{3}\gamma_2 k_xk_y\right]$, 
$S_k = \frac{\hbar^2\gamma_3}{2m_0}\sqrt{3}\left(k_x-ik_y\right)k_z$.
The matrix of the strain part of the Hamiltonian, $H_{\epsilon}^h$, is in the same form of Eq.(\ref{hammat}) but with the replacements of the operators $\{ P_{k}$, $R_{k}$, $S_{k}$, $R_{k}\}$ by $\{ P_{\epsilon}$, $R_{\epsilon}$, $S_{\epsilon}$, $R_{\epsilon}\}$ that are generated by the rules of transformation: $k_{\alpha}k_{\beta} \rightarrow \epsilon_{\alpha\beta}$, $\frac{\hbar^2 \gamma_1}{2m_0} \rightarrow -a_v$, $\frac{\hbar^2 \gamma_2}{2m_0} \rightarrow -\frac{b}{2}$, $\frac{\hbar^2 \gamma_3}{2m_0} \rightarrow -\frac{d}{2\sqrt{3}}$. \cite{Chuang}
The parameters $\gamma_1=7.1$, $\gamma_2=2.02$, $\gamma_3=2.91$ ,$a_v=1.16eV$, $b=-1.7eV$, and $d=-4.55eV$ are taken for a valence hole in a stressed GaAs material.\cite{Bimberg,Chuang}

\begin{figure}[t]
\includegraphics[width=15.0cm]{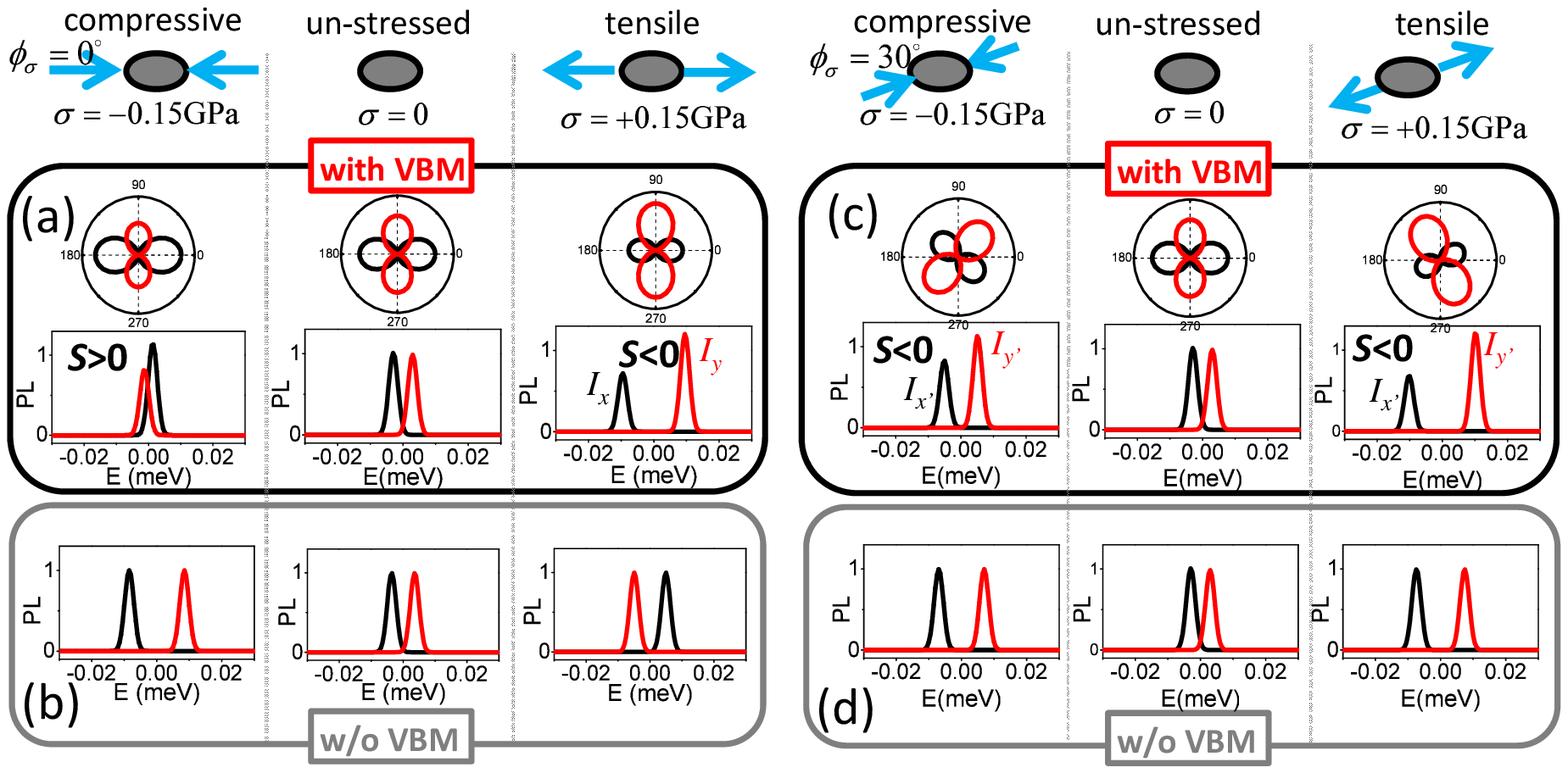}
\caption{(a) [(c)] Numerically calculated polarized emission spectra and the corresponding polar plots of a $x$-elongated QD under uni-axial stresses with $\phi_{\sigma}=0^\circ$ [$\phi_{\sigma}=30^\circ$] and strength $\sigma=0,\pm 0.15$GPa. (c) [(d)]: same as (a) [(b)] but without the consideration of VBM.  }
\label{fig2}
\end{figure}

\begin{figure}[b]
\includegraphics[width=12.0cm]{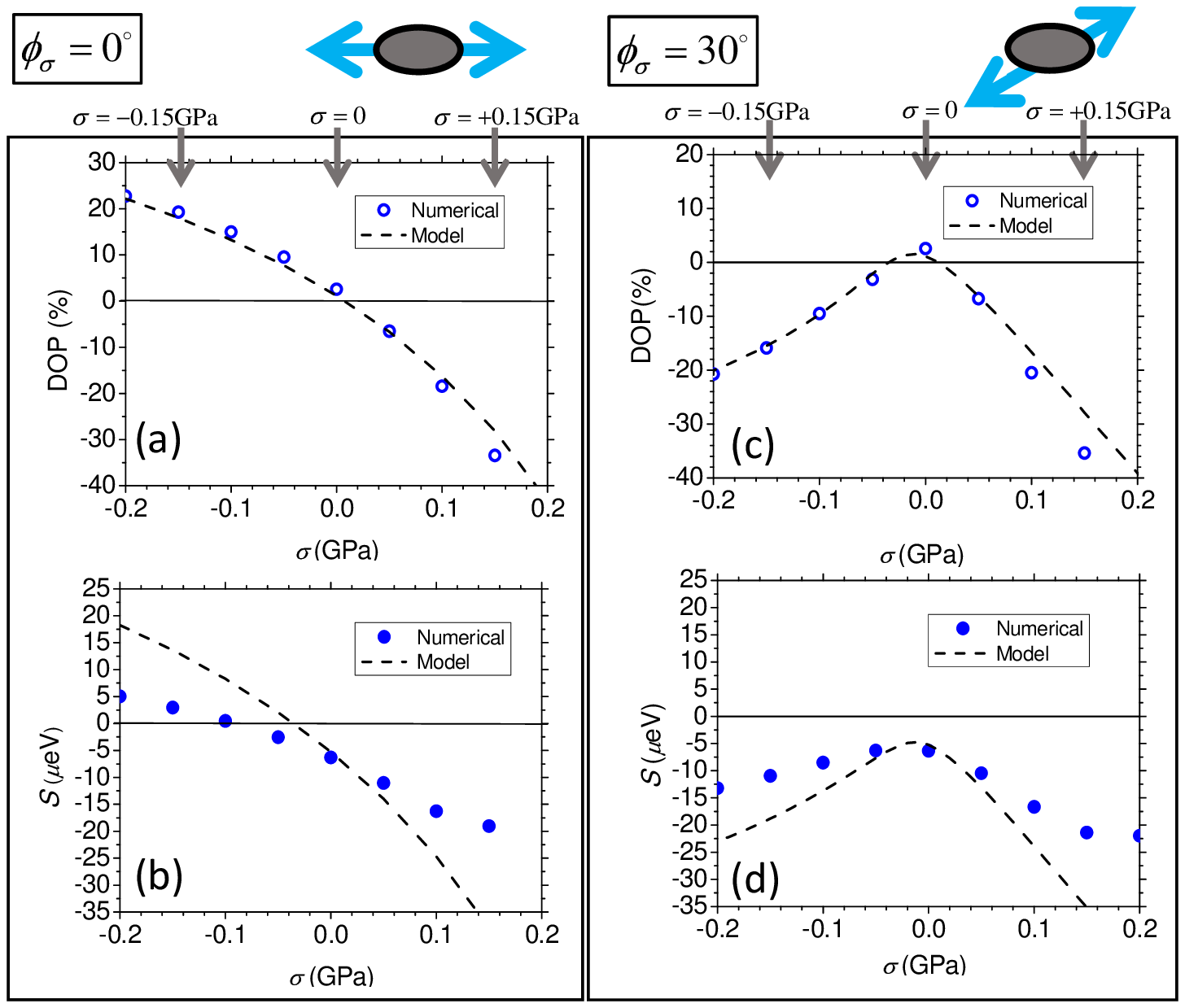}
\caption{ (a) The excitonic fine structure splitting, $S$, defined by Eq.(\ref{Sdef}) and (b) the degree of polarization, DOP, defined by Eq.(\ref{DOPdef}) of the bright exciton doublet of an $x$-elongated QD under an uni-axial stress of varied strength $\sigma$ along the $x$-axis as considered for Fig.~\ref{fig2}(a). Solid (Dotted) lines: numerically calculated results obtained by $k \cdot p$ theory (results yielded by the formalisms based on the model presented in Sec.\ref{paramodel}). (c) and (d): the numerical and analytical results for the same dot but with the uniaxial stress misaligned to the $x$-axis with $\phi_{\sigma}=30^\circ$ as considered for Fig.~\ref{fig2}(c). }
\label{fig3}
\end{figure}

\section{Numerical computations}

\subsection{Methods}
\subsubsection{Single particle spectra}
Numerically, the single-electron (-hole) energy spectrum $E_{i_e}^e$  ($E_{i_h}^h$) and the envelope functions $g_{s_z}^{e}$ ($\{g_{j_z}^{h}\}$) are calculated using finite difference method for a GaAs/Al$_{0.35}$Ga$_{0.65}$As DE-QDs shaped by a Gaussian profile $0 \le z \le H \exp(-\frac{x^2}{\Lambda_x^2}-\frac{y^2}{\Lambda_y^2})$, characterized by the height of QD $(H)$ and the parameters of lateral characteristic length of QD along the $x/y$ direction  $(\Lambda_{x/y})$.\cite{Liao}  Accordingly, one can define the characteristic function for a QD, $X_{QD}(\vec{r})$, that is equal to one (zero) as the coordinate position $\vec{r}$ in (out of) the QD.  
Thus, the confining potential of a GaAs/Al$_{0.35}$Ga$_{0.75}$As QD for an electron (a hole) can be expressed as $V_{QD}^{e/h}(\vec{r}_{e/h})=V_b^{e/h} \cdot X_{QD}(\vec{r}_{e/h})$ with the band-offset $V_b^e=300$meV ($V_b^h=200$meV). \cite{Chuang}.  
Throughout this work, we choose the Cartesian coordinate frame with the axes along the specified crystalline directions, i.e. $\hat x \parallel [1\bar{1}0]$, $\hat y \parallel [110]$ and $\hat z \parallel [001]$, and study the asymmetric QDs that are elongated along the $x$-axis with $\Lambda_{x}>\Lambda_{y}$, as depicted in Fig.\ref{fig1}. Besides, a QD might be considered to be under an uniaxial stress of strength $\sigma$ in the direction of $\hat{n}_{\sigma}=(\cos\phi_{\sigma},\sin\phi_{\sigma},0)$, as depicted in Figs.~\ref{fig1}(a) and (b), which yields the strain tensor elements given by 
$\epsilon_{xx} = \left( \frac{s_{11}+s_{12}}{2}\right) \sigma + \left( \frac{s_{44}}{2}\right) \sigma \cos 2\phi_{\sigma}$, $\epsilon_{yy} = \left( \frac{s_{11}+s_{12}}{2}\right) \sigma - \left( \frac{s_{44}}{2}\right) \sigma \cos 2\phi_{\sigma}$, $\epsilon_{zz} = s_{12} \sigma$, $\epsilon_{xy} = \left( \frac{s_{11}-s_{12}}{2}\right) \sigma \sin 2\phi_{\sigma}$,
where the elastic compliance constants are $s_{11}=0.0082$GPa$^{-1}$, $s_{12}=-0.002$GPa$^{-1}$, and $s_{44}=0.0168$GPa$^{-1}$ for GaAs.\cite{Chuang}

\subsubsection{Electron-hole exchange interactions}
Since our interest is in the fine structures of the lowest exciton doublet, we take into account only the relevant lowest single-electron and single-hole orbitals and, for brevity of notation, label them simply with the spin indices, i.e. $\psi^e_{i_e=\uparrow_e/\downarrow_e} = |\uparrow_e/\downarrow_e \rangle $, ($\psi^h_{i_h=\Uparrow_h'/\Downarrow_h'} = |\Uparrow_h'/\Downarrow_h' \rangle$). Here, a superscript prime is attached to the arrow symbol of hole spin to indicate the VBM nature that mixes the heavy-hole (HH) and light-hole (LH) components.
In the basis of the direct products of the single-electron and -hole states, $|\uparrow_e\rangle |\Downarrow_h'\rangle$ and $|\downarrow_e \rangle |\Uparrow_h'\rangle$, being the two bright exciton (BX) configurations, the Hamiltonian for an VBM BX in a QD is written as a $2\times 2$ matrix,  
\begin{eqnarray}\label{hamilX}
H_X= 
\left(
       \begin{array}{cc}
E_X^{(0)} & \tilde{\Delta}_{eff}^{xc} \\
\tilde{\Delta}_{eff}^{xc \, \ast} & E_X^{(0)} \\
       \end{array}
     \right) \, ,
\end{eqnarray}
where $E_X^{(0)}=E_{\uparrow_e}^e+E_{\Downarrow_h'}^h -V_{\uparrow_e\Downarrow_h'\Downarrow_h'\uparrow_e}^{eh} = E_{\downarrow_e}^e+E_{\Uparrow_h'}^h -V_{\downarrow_e\Uparrow_h'\Uparrow_h'\downarrow_e}^{eh}$ denotes the energy of exciton regardless of the $\it e-h$ exchange interactions, and $\tilde{\Delta}_{eff}^{xc}\equiv V_{\uparrow \Downarrow' \Uparrow' \downarrow }^{ehxc}$ is the off-diagonal matrix element of $\it e-h$ exchange interaction that couples the two VBM bright exciton (BX) configurations of opposite angular momenta and results in the FSS of the exciton doublet, $|S|=2|\tilde{\Delta}_{eff}^{xc}|$. One should note that the off-diagonal matrix element in general is complex, and can be written as
\begin{equation}\label{Delta_eff_1_2}
\tilde{\Delta}_{eff}^{xc} = \Delta_{eff,1}^{xc} +i \Delta_{eff,2}^{xc} \equiv \Delta_{eff}^{xc} e^{-i\theta_{eff}}
\end{equation}
where $\Delta_{eff,1}^{xc}$ ($\Delta_{eff,2}^{xc}$) is the real (imaginary) part and $\Delta_{eff}^{xc}$ is the magnitude. As will be shown later, the phase angle $\theta_{eff}$ is essential in the determination of the orientations of the optical polarizations, the pseudo-spins, and the Bloch vectors of the resulting exciton eigen states.

In the numerical calculation, the matrix elements of {\it e-h} exchange interactions are divided by the short-ranged and long-ranged parts according to the averaged Wigner-Seitz radius, and computed separately.\cite{Kadantsev} The former is treated in the dipole-dipole interaction approximation and numerically integrated using trapezoidal rules and graphics processing unit (GPU) parallel computing technique for numerical acceleration.  The latter is considered for the matrix elements involving the exciton basis of same angular momenta and evaluated using the formalism of Eq.(2.17) in Ref.\cite{Taka}, in terms of the the energy splitting between the bright- (BX) and dark-exciton (DX) states of a QD, $E_X^S = \Delta_{eh,bulk}^{xc} \times [ \pi (a_B^\ast)^3 \int d^3r |g_{s_z=\pm\frac{1}{2}}^e|^2 |g_{j_z=\mp\frac{3}{2}}^h|^2 ]$ (See also Eq.(2) in Ref.\cite{Liao}), which is extrapolated from the the BX-DX splitting $\Delta_{eh,bulk}^{xc}=20\mu$eV of a pure HH-exciton with the effective Bohr radius $a_B^\ast$ in GaAs bulk. \cite{Ekardt} 
 While the screening in the e-h exchange interactions is known as a subtle problem,
here we follow Ref.\cite{Allan} and assume that the long ranged electron-hole exchange interactions are screened by the static dielectric constant $\epsilon_b=12.9$ of host material GaAs. By contrast, we consider that the screening in the short ranged interactions has been implicitly merged in the empirical parameter of $\Delta_{eh,bulk}^{xc}$, as discussed by Kadantsev and Hawrylak \cite{Kadantsev}, and do not include the background dielectric constant in the formalism.  

\subsection{Stress-dependent polarized fine structures}

By solving the Schr\"odinger equation, $H_X |\Psi_n^{X}\rangle =E_n^X |\Psi_n^{X}\rangle $, we obtain the eigen states and the energies of the BX doublet, $E_{\pm}^X=E_X^{(0)} \pm \Delta_{eff}^{xc}$. Accordingly, the intensity of the emitted light, $I_n(\hat{e},\omega) \propto |\langle 0|P_{\hat{e}}^- | \Psi_{n}^X \rangle|^2 \delta(\hbar \omega -E_n^X)$, of the  frequency $\omega$ and polarization along the $\hat{e}$-direction  from a stressed QD can be calculated using the formalism of Fermi's golden rule,\cite{Kumar} where $P_{\hat{e}}^-$ is the polarization operator as defined in Eqs.(18) and (19) in Ref.\cite{Liao}. For the maximum intensity of the emitted light from an exciton state $\Psi_n^X$ that is polarized along the optical axis, $\hat{e}_0$, we simplify its notation as $I_{n}(\hat{e}_0;\omega=E_n^X/\hbar) \equiv I_{e_0}$.

In the presence of stress, the optical axes of the exciton states of a QD might be re-directed and not any more aligned to the $x$- or $y$-axes. Here, we specify the re-directed optical axis of a stressed QD that is directionally close to the $x$-axis ($y$-axis) as the $x'$-axis ($y'$-axis) (See Fig.~\ref{fig1}(c)).
To characterize a polarized fine structure of a stressed QD, the parameter of degree of polarization,
\begin{equation}\label{DOPdef}
\text{DOP} \equiv \frac{ I_{x'}-I_{y'}}{I_{x'}+I_{y'}} \, ,
\end{equation}
and that of fine structure splitting, 
\begin{equation}\label{Sdef}
S \equiv E_{x'}^X-E_{y'}^X \, 
\end{equation}
are defined.
Here, the subscript $x'$ ($y'$) indicates the direction of the optical axis of an exciton state, $|\Psi_{x'}^X\rangle $ ($|\Psi_{y'}^X\rangle $), and is also used to label the corresponding energy and emission intensity.
Note that the signs of the defined DOP and $S$ depend on the relative intensities and the order of the energies of the emission lines in the FS.

\subsubsection{Aligned stresses}
Figures \ref{fig2} (a) shows the numerically calculated polarized emission spectra of the $x-$elongated GaAs/AlGaAs QD  of $\Lambda_x=14$nm, $\Lambda_y=12.7$nm and $H=9$nm, applied by uniaxial stresses, aligned to the $x$-axis, of different strengths and types of $\sigma=0$ (stress-free), $-0.15$GPa (compressive) and $+0.15$GPa (tensile), respectively. 
In the stress-free case, the FS spectrum of the $x$-elongated DE-QD is featured by the low-energy $x$-polarized and the high energy $y$-polarized lines that are split by $|S|\sim 7\mu$eV and slightly differing in the intensities, characterized by DOP$\sim 3 \%$. The $x$-polarized emission line lying at lower energy in the excitonic FS of the QD results from the dominance of the long ranged  {\it e-h} exchange interactions in the large QD that are essentially dipole-dipole interactions and energetically favour the exciton state that is optically $x$-polarized, along the elongation axis of QD. \cite{Lin} 

Applying an $x$-aligned stress onto a QD substantially affects the FS feature described above. In Fig.~\ref{fig2} (a), we observe the obvious changes of the magnitude of the FSS of the QD caused by applying the stresses of $\sigma=-0.15$GPa and $+0.15$GPa. Remarkably, the types (compressive or tensile) of stress also affect the order of the $x$- and $y$-polarized emission lines of the stressed QD.

Figure ~\ref{fig3} (a) and (b) present the numerically calculated DOP and $S$ of the stressed QD of Fig.\ref{fig2} (a), respectively, against the $x$-aligned uniaxial stress of the strength continuously varied from $\sigma=-0.2$GPa to $\sigma=+0.2$GPa.  More clearly, it is shown that overall the magnitudes of the $S$ and DOP are increased by increasing the strength of the applied stress with $|\sigma| > 0.1$GPa, but the signs of the $S$ and DOP change from positive to negative as the applied compressive stress is changed to be tensile. The sign change of the S and DOP reflects the reversal of the order of the $x$- and $y$-polarized emission lines in energy.

As known from previous studies, the stress-dependent DOP of a stressed QD is associated with the stress-enhanced VBM, so is the FSS.\cite{Santosh} 
To highlight the VBM effect, Fig. \ref{fig2} (b) presents the emission spectra of the same stressed QD that are calculated regardless of the VBM (by artificially setting $S_k=S_\epsilon=0$ and $R_k=R-\epsilon=0$ in the $k\cdot p$ Hamiltonian), showing completely different features from that of Fig. \ref{fig2} (a). More detailed analysis of the stress-induced VBM effects in Fig. \ref{fig2} will be presented in the next section.

\subsubsection{Misaligned stresses}
Next, let us consider the applied uniaxial stress misaligned to the elongation axis (the $x$-axis) of QD. Figure \ref{fig2}(c) shows the numerically calculated polarized emission spectra of the $x$-elongated QD under the misaligned uni-axial stresses of magnitudes $\sigma=0,\pm 0.15$GPa that are counter-clockwise rotated from and misaligned to the $x$-axis by $\phi_{\sigma}=30^\circ$.   Figure \ref{fig3} (c) and (d) present the calculated DOP and $S$ of the QD under the misaligned uniaxial stresses as function of the magnitude of the stress. As compared with the cases of aligned stress ($\phi_{\sigma}=0^\circ$), the FSS's of the QD with misaligned stresses are shown always non-vanishing, with a lower bound of $|S|\sim 6\mu eV$ at $\sigma =-0.02$GPa, as observed and predicted by Refs.~\cite{Plumhof,Santosh,Singh}. Another obvious observation is that the optical polarization axes of the QD under the misaligned uniaxial stresses are no longer aligned to either the $x$- or the $y$-axes, but directed in between and accompanied with significant changes of the magnitudes of the FSS and DOP. As will be elucidated more in the analysis of the next section, the rotation of the optical axes is understood as a resulting optical feature from  the superposition of exciton eigen states of the stressed QD, mixed with the stress-free $x-$ and $y-$ polarized exciton states by a misaligned stress.

\section{Model analysis}

\subsection{Pseudo-spin representation}

To elucidate the effects of uniaxial stress, we take the optically $x-$ and $y-$ polarized exciton configuration, $\frac{1}{\sqrt{2}}\left(|\downarrow_e \rangle |\Uparrow_h'\rangle +  |\uparrow_e\rangle |\Downarrow_h'\rangle \right) \equiv |\Psi_x^X\rangle $ and $ \frac{-i}{\sqrt{2}} \left(|\downarrow_e \rangle |\Uparrow_h'\rangle -  |\uparrow_e\rangle |\Downarrow_h'\rangle \right) \equiv |\Psi_y^X \rangle $,  as basis for expanding the undetermined exciton states and constructing an effective Hamiltonian matrix of a QD with an uniaxial stress along an arbitrary direction. 
In the chosen basis $\{ |\Psi_x^X\rangle, |\Psi_y^X\rangle \} $, the $2 \times 2$ matrix of Hamiltonian for a VBM exciton in a stressed QD can be expressed in a compact form as  
\begin{equation}\label{Blocheq}
H_{X}' = \vec{\sigma} \cdot \vec{\Omega}_{eff} \, ,
\end{equation}
where $H_{X}'=H_{X} - E_{X}^{(0)}$ is the exciton Hamiltonian offset by the spin-independent averaged energy of BX doublet, $\vec{\sigma}=(\sigma_1, \sigma_2, \sigma_3)$ is the vector with the components of Pauli-matrices,
\begin{eqnarray}
\sigma_1= 
\left(
       \begin{array}{cc}
0 & 1 \\
1 & 0 \\
       \end{array}
     \right) \, ,
\sigma_2= 
\left(
       \begin{array}{cc}
0 & -i \\
i & 0 \\
       \end{array}
     \right) \, , 
\sigma_3= 
\left(
       \begin{array}{cc}
1 & 0 \\
0 & -1 \\
       \end{array}
     \right) \, ,    
\end{eqnarray}
and  
\begin{equation}\label{effectfield}
\vec{\Omega}_{eff} \equiv  (\Omega_{1},\Omega_{2},\Omega_{3}) = \Delta_{eff}^{xc}(\sin\theta_{eff}, 0, \cos\theta_{eff}) 
\end{equation}
acts as a pseudo-magnetic field that is coupled to the pseudo-spin of the exciton doublet represented by $\vec{\sigma}$ and orientated to the direction of $(\sin\theta_{eff}, 0, \cos\theta_{eff})$.

\begin{figure}[b]
\includegraphics[width=12.0cm]{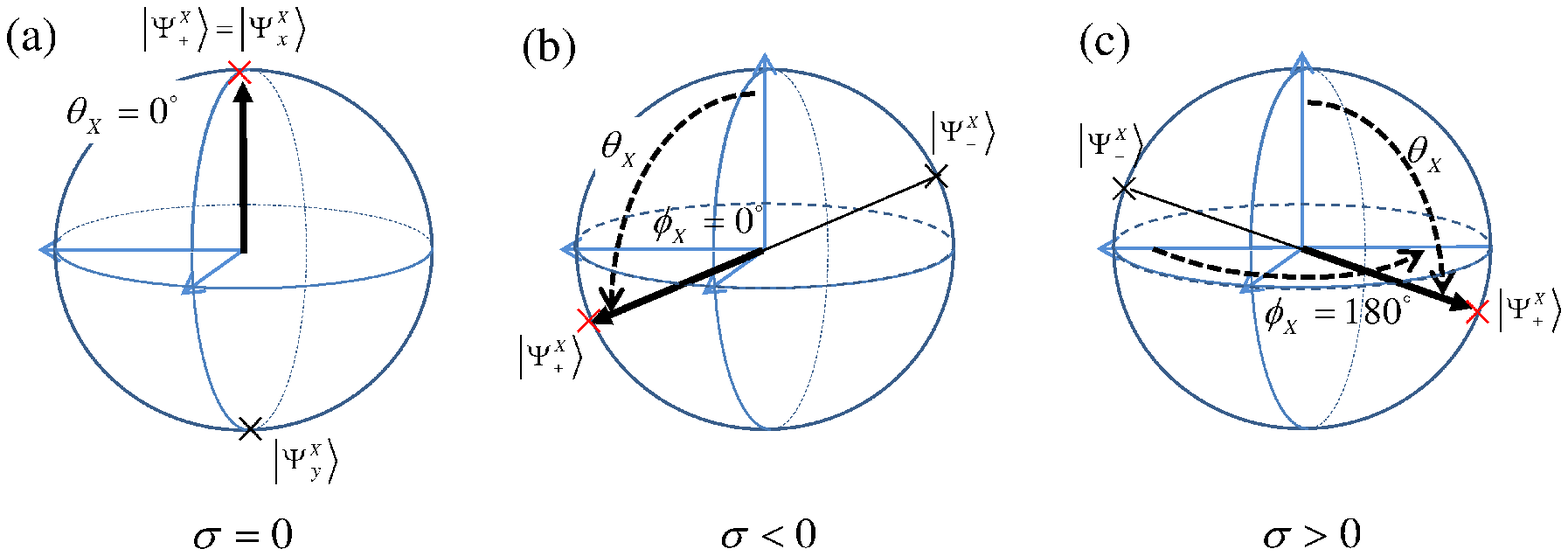}
\caption{Schematics of Bloch vectors of exciton eigen states of stress-free and stressed QDs. (a) As an example, the high energy (HE) exciton eigen states $\Psi_{+}^X = \Psi_{x}^X$ are $x$-polarized and geometrically specified to the north poles of the Bloch sphere. As a counterpart, the low-energy (LE) $y$-polarized state $\Psi_{-}^X = \Psi_{y}^X$ is at the south pole. (b) [(c)] Applying an compressive [tensile] uniaxial stress to the QD makes the exciton eigen states mixed by $\Psi_{x}^X$ and $\Psi_{y}^X$, the HE one of which, $\Psi_{+}^X$, is represented by a Bloch vector positioned between the north- and south-poles on the longitude of $\phi_X=0^\circ$ [$\phi_X=180^\circ$].}
\label{fig4}
\end{figure}

Next, by solving Eq.(\ref{Blocheq}) we obtain that the energies of exciton eigen states $|\Psi_{\pm}^X\rangle $ are given by $E_{\pm}^X=E_{X}^{(0)}\pm {\Delta}_{eff}^{xc}$, split by the FSS $|S|=2\Delta_{eff}^{xc}=2|\vec{\Omega}_{eff}|$.
In the generalized Bloch vector representation, the high energy (HE) and the low energy (LE) exciton eigen states can be expressed as 
\begin{eqnarray}
|\Psi_{+}^{X}\rangle &=& (\cos\frac{\theta_{X}}{2}, e^{-i\phi_{X}}\sin\frac{\theta_{X}}{2}) \, , \label{Blochv1}\\
|\Psi_{-}^{X}\rangle &=& (\sin\frac{\theta_{X}}{2}, -e^{-i\phi_{X}}\cos\frac{\theta_{X}}{2}) \, , \label{Blochv2}
\end{eqnarray} 
respectively, characterized by the phase angles $\theta_X$ and $\phi_X$. 
Comparing Eqs. (\ref{Blochv1}) and (\ref{Blochv2}) with the solved exciton eigen states from Eqs.(\ref{Blocheq}-\ref{effectfield}), one can relate the phase angles of Bloch vector to $\theta_{eff}$:
\begin{equation}
\theta_X=\theta_{eff}\, , \phi_X=0^\circ \, \text{ as } \theta_{eff}=\{0^\circ, 180^\circ\} \label{cond1}
\end{equation}
and
\begin{equation}
\theta_X=360^\circ - \theta_{eff}\, , \phi_X=180^\circ \, \text{ as } \theta_{eff}=\{180^\circ, 360^\circ\} \, . \label{cond2}
\end{equation}

\subsection{3D parabolic Model}\label{paramodel}

For more analysis, we take the three-dimensional (3D) parabolic model for the confining potential of DE-QD, \cite{Liao} yielding the solvable single particle wave functions and energy spectra.  
Within the model, the envelope wave function of the lowest single-electron state of a QD is explicitly given by $\phi_{000}^{e} =(\frac{1}{{\pi^{\frac{3}{2}}l_{x}^{e} l_{y}^{e} l_{z}^{e}}})^{1/2} \exp{\{-\frac{1}{2}[(\frac{x}{l_{x}^{e}})^{2}+(\frac{y}{l_{y}^{e}})^{2}+(\frac{z}{l_{z}^{e}})^{2}]\}}$, in terms of the parameters, $l_{x}^{e}$, $l_{y}^{e}$, and $l_{z}^{e}$, of the wave function extents in the $x$-, $y$-, and $z$-directions, respectively.  
By contrast, the energy spectrum and wave function of a single hole in a QD are hardly solved analytically even within the simplified parabolic model because of the off-diagonal elements in the Hamiltonian of Eq.(\ref{hammat}) that make the mixture of HH and LH components, i.e. the VBM. Regardless of the VBM (by setting $S_k=S_\epsilon=0$ and $R_k=R_\epsilon=0$ in Eq.(\ref{hammat})), the eigen states of a pure HH or LH in QD in the parabolic model can be described by the same formula of wave function as that of electron, $\phi_{000}^{HH/LH}$, with the substitution of the length parameters for a HH or a LH, $l_{x,y,z}^{HH/LH}$. Throughout this work, we consider $x$-elongated QDs with $\eta\equiv l_x/l_y \ge 1$. For brevity of notation, hereafter we denote the lowest pure HH (LH) states as  $|\psi_{\Uparrow/\Downarrow}^h\rangle \equiv |\Uparrow_h/\Downarrow_h \rangle \approx |\phi_{000}^{HH} u_{+\frac{3}{2}/-\frac{3}{2}}^h \rangle $ ($|\psi_{\uparrow/\downarrow}^h\rangle \equiv |\uparrow_h/\downarrow_h \rangle \approx |\phi_{000}^{HH} u_{+\frac{1}{2}/-\frac{1}{2}}^h \rangle $). 
 
Following Ref.\cite{Leger}, one can treat the HH-LH coupling terms ($R_k,R_\epsilon,S_k$, and $S_\epsilon$) in the hole Hamiltonian as perturbations and, in the lowest order approximation, write the expanded VBM hole states of a QD as   
\begin{eqnarray}\label{VBM-state}
|\Uparrow_h'\rangle &\approx & |\Uparrow_h \rangle  - \tilde{\beta}_{HL}^\ast |\downarrow_h\rangle \, , \nonumber \\
|\Downarrow_h'\rangle &\approx & |\Downarrow_h \rangle  - \tilde{\beta}_{HL} |\uparrow_h\rangle \, ,
\end{eqnarray}
where the (complex) coefficient for the most leading LH component is
\begin{equation}\label{betahl}
\tilde{\beta}_{HL}= \frac{\tilde{\rho}_{HL,k} + \tilde{\rho}_{HL,\epsilon}}{\Delta_{HL}} \, ,
\end{equation}
determined by the ratio of the matrix element of the HH-LH coupling operator $\hat{R}= \hat{R}_k + \hat{R}_{\epsilon} $, $\tilde{\rho}_{HL,k(\epsilon)} \equiv \langle \phi_{000}^{HH} |\hat{R}_{k(\epsilon)} | \phi_{000}^{LH} \rangle $, and the energy difference between the HH- and LH-levels, $\Delta_{HL} \equiv  \langle \phi_{000}^{LH} |\hat{P}-\hat{Q}+V_{QD} | \phi_{000}^{LH} \rangle - \langle \phi_{000}^{HH} |\hat{P}+\hat{Q}+V_{QD} | \phi_{000}^{HH} \rangle$, where $\hat{P}\equiv \hat{P}_k + \hat{P}_{\epsilon}$ and $\hat{Q}\equiv \hat{Q}_k + \hat{Q}_{\epsilon}$.\cite{Lin}  
Taking into account Eq.(\ref{VBM-state}), the matrix element of the {\it e-h} exchange interaction for a VBM-exciton is derived as $\tilde{\Delta}_{eff}^{xc} \equiv V_{\uparrow \Downarrow' \Uparrow' \downarrow }^{eh,xc} \approx V_{\uparrow \Downarrow \Uparrow \downarrow }^{eh,xc} -2 \tilde{\beta}_{HL} V_{\uparrow_e \Downarrow_h \downarrow_h \downarrow_e}^{eh,xc}$, or equivalently expressed, in a similar form presented in Ref.\cite{Lin}, as 
\begin{equation}\label{Delta_eff}
\tilde{\Delta}_{eff}^{xc}= -\Delta_1 + \tilde{\Delta}_{VBM} \, ,
\end{equation}
where the first term $-\Delta_1\equiv   V_{\downarrow_e \Uparrow_h \Downarrow_h \uparrow_e}^{eh,xc}$ is the matrix element of the long ranged {\it e-h} exchange interaction that couples the two pure-HH exciton configurations of opposite angular momenta, $|\downarrow_e \Uparrow_h \rangle$ and $|\uparrow_e \Downarrow_h \rangle$, and the second term is
\begin{equation}\label{DeltaVBM}
 \tilde{\Delta}_{VBM} =  \frac{2E_X^S}{\sqrt{3}} \tilde{\beta}_{HL}\, ,
\end{equation}
that originates from the (short ranged) interaction $E_S^X \equiv V_{\downarrow_e \Uparrow_h \Uparrow_h \downarrow_e }^{eh,xc} (= -\sqrt{3}V_{\downarrow_e \Uparrow_h \uparrow_h \uparrow_e }^{eh,xc}=V_{\uparrow_e \Downarrow_h \Downarrow_h \uparrow_e }^{eh,xc}= -\sqrt{3} V_{\uparrow_e \Downarrow_h \downarrow_h \downarrow_e }^{eh,xc})$ that makes the BX-DX splitting and is involved in the FSS of an exciton via VBM.
The r.h.s of Eq.(\ref{Delta_eff}) is formulated in such a way to stress the attractive nature of the long ranged interaction ($-\Delta_1$) with respect to the $x$-polarized exciton FSS state and the repulsive VBM-induced interaction $(\tilde{\Delta}_{VBM})$, which might energetically compensates or even overwhelms the attractive interaction $(-\Delta_1)$.

In the parabolic model, one can derive all the terms used in Eqs.(\ref{betahl})-(\ref{DeltaVBM}) explicitly in terms of the QD and material parameters, which are
\begin{equation}\label{rhok}
\tilde{\rho}_{HL,k}=\rho_{HL,k}  = \frac{\sqrt{3}\hbar^2\gamma_3}{4m_0}\cdot \left[ \left( \frac{1}{l_y^h} \right)^2 - \left( \frac{1}{l_x^h} \right)^2 \right] \, ,
\end{equation}
\begin{equation}\label{rhoepsilon}
\tilde{\rho}_{HL,\epsilon}  =  -   \frac{|d|  s_{44}}{4}   \sigma \cos 2\phi_{\sigma} + i  \frac{\sqrt{3}|b| (s_{11}-s_{12})}{2} \sigma \sin 2\phi_{\sigma} \, ,
\end{equation}
$\Delta_{HL} \approx \frac{\hbar^2\gamma_2}{m_0}\cdot \frac{1}{l_z^2} -|b|(s_{11}-s_{12})\sigma $,  $E_X^S \approx  \dfrac{1}{2}\times\dfrac{{a_B^{\ast}}^3\Delta_{eh,{\rm bulk}}^{xc}}{\sqrt{8\pi}l_x^{eh}l_y^{eh}l_z^{eh}}$ \cite{Taka,Lin}, and $\Delta_1=\frac{1}{4 \pi \varepsilon_0}\frac{3 \sqrt{\pi} e^2 \hbar^2 E_p}{16\sqrt {2} m_0 {E_g}^2} \frac{\eta(\eta-1)}{(l_x^{eh})^3}e^{(\frac{3 \sqrt{\pi} l_z^{eh}}{4 l_y^{eh}})^{2}} {\rm erfc}(\frac{3 \sqrt{\pi} l_z^{eh}}{4 l_y^{eh}})$ \cite{Ramirez}, where $a_B^\ast= 11nm$ ($\Delta_{eh,{\rm bulk}}^{xc}=20\mu \rm{eV}$) is the effective Bohr radius (the BX- and DX-level splitting) of exciton in bulk GaAs, $E_P=28.8eV$, and $E_{g}=1.519eV$ is the energy gap of GaAs.\cite{Bimberg,Chuang}  Here, $l_{x,y,z}^{HH}=l_{x,y,z}^{LH}=l_{x,y,z}^h$ and $l_{\alpha}^{eh} \equiv \frac{\sqrt{2}l_{\alpha}^e l_{\alpha}^h}{\sqrt{(l_{\alpha}^e)^2 (l_{\alpha}^h)^2}}$ are assumed for the compactness of formalisms.\cite{Ramirez} 
From Eqs.(\ref{Delta_eff})-(\ref{rhoepsilon}), the magnitude of FSS defined by Eq.(\ref{Sdef}) is given by $|S|=2|\tilde{\Delta}_{eff}^{xc}|=2|\Delta_{eff,1}^{xc} + i\Delta_{eff,2}^{xc} |$, where
\begin{equation}\label{Deltaeff1}
\Delta_{eff,1}^{xc}= -\Delta_1 + \frac{2E_X^S}{\sqrt{3}\Delta_{HL}} (\rho_{HL,k} -   \frac{|d|  s_{44}}{4}   \sigma \cos 2\phi_{\sigma})
\end{equation}
and 
\begin{equation}\label{Deltaeff2}
\Delta_{eff,2}^{xc} = \frac{2E_X^S}{\sqrt{3}\Delta_{HL}} \frac{\sqrt{3}|b| (s_{11}-s_{12})}{2} \sigma \sin 2\phi_{\sigma} \, .
\end{equation}
Notably, only a misaligned stress ($\sigma\neq 0, \phi_{\sigma}\neq 0,\pi$) can yield a non-vanishing imaginary part, $\Delta_{eff,2}^{xc}$, and, according to Eq.(\ref{Delta_eff_1_2}), gives rise to a phase angle $\theta_{eff}\neq  0$.

It is indicated from Eq.(\ref{Deltaeff2}) that $\Delta_{eff,2}^{xc}<0$ ($\Delta_{eff,2}^{xc}>0$) and, according to Eq.(\ref{Delta_eff_1_2}) , the resulting phase angle $\theta_{eff}$ falls into the range, $\theta_{eff}=\{0^\circ, 180^\circ \}$ ($\theta_{eff}=\{180^\circ, 360^\circ \}$), as a compressive (tensile) uniaxial stress with $0<\phi_{\sigma}<45^\circ$ is applied to a QD.  Accordingly, Eq.(\ref{cond1}) can be used to determine the possible range of the orientation of an exciton Bloch vector on the Bloch sphere for QDs with compressive stress while Eq.(\ref{cond2}) is for QDs with tensile stress. 
 Figure ~\ref{fig4} (a),(b) and (c) depict the Bloch vectors of the HE exciton states $|\Psi_{+}^{X}\rangle$ of stress-free, compressively, and tensile stressed QDs on the Bloch spheres, respectively.

Using the formalism of the Fermi's golden rule in Ref.\cite{Liao} and Eqs.(\ref{Delta_eff_1_2}),(\ref{Blochv1})-(\ref{cond2}) and (\ref{VBM-state})-(\ref{DeltaVBM}), one can derive the intensities of the $\hat{e}$-polarized ($\hat{e}=(\cos\phi, \sin\phi, 0)$) emitted lights from the exciton eigen states, $|\Psi_{\pm}^X\rangle $, of a uniaxially stressed QD as, 
\begin{equation}\label{optaxes1} 
I_{+}(\phi)  \propto  I_{+,max} \cos^2(\phi-\phi_{+}) 
\end{equation}
and
\begin{equation}\label{optaxes2} 
I_{-}(\phi)  \propto  I_{-,max} \sin^2(\phi-\phi_{-}) \, ,
\end{equation}
respectively, where the maximum intensities are determined by 
\begin{equation}\label{intensitymax}
I_{\pm,max}=  (1\pm \frac{\Delta_{eff}^{xc} +\Delta_1 \cos\theta_{eff}}{2E_{X}^S})^2 +(\frac{\Delta_1 \sin\theta_{eff}}{2E_{X}^S})^2\, , 
\end{equation}
and the angle of the optical axis for $I_{+}$ ($I_{-}$) with respective to the $x$-  ($y$)-axis is
\begin{equation} \label{phipm}
\phi_{\pm} = \frac{\theta_{eff}}{2} + \delta\phi_{\pm}\sim \frac{\theta_{eff}}{2}\, ,
\end{equation}
where $\delta\phi_{\pm}=\tan^{-1}\left( \frac{\mp \Delta_1 \sin \theta_{eff}}{2E_{X}^S \pm \Delta_{eff} \pm \Delta_1 \cos\theta_{eff} } \right)$. Thus, the magnitude of the DOP of the emission lines from the exciton doublet is given by 
$|{\rm{DOP}}| = \left|\frac{I_{+,max}- I_{-,max}}{I_{+,max} + I_{-,max}} \right|$.
Equation (\ref{phipm}) shows that the orientation of the optical polarization of an exciton state in the FS of a QD is along the direction rotated from the $x$- or $y$-axes by the angle $\sim \theta_{eff}/2 $, which is specified by the new $x'$- or $y'$-axes as depicted in Fig.~\ref{fig1}.  
Using the above simplified model, the S's and DOP's of the stressed QD considered in Figs.~\ref{fig2} and \ref{fig3} are calculated \cite{parameters} and  show qualitative agreements with the numerical results, as seen in Fig.\ref{fig3}.  

\begin{figure}[t]
\includegraphics[width=12.0cm]{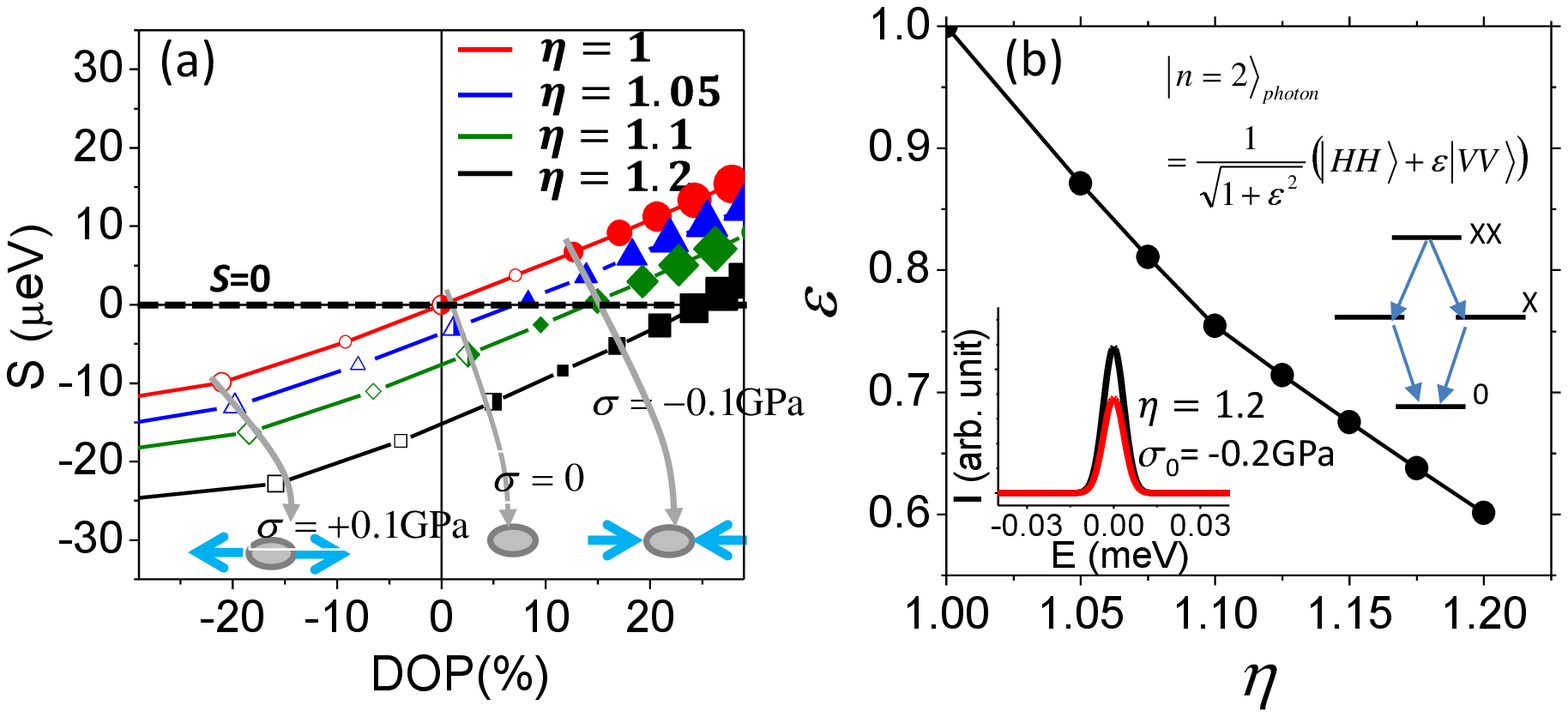}
\caption{(a) $S$ versus DOP of the polarized emission lines from the bright exciton doublets of the $x$-elongated QDs with $\eta\equiv l_x/l_y =1, 1.05, 1.1, 1.2$ under uniaxial stresses of $\sigma=0.1,0.05,...,-0.3$GPa along the elongation axis of QD. The areas of the empty (filled) symbols reflect the magnitudes of the applied tensile (compressive) stresses. Note that the resulting DOP's ($\neq 0$) of the stressed QDs with $S=0$ are non-zero and lead to the non-maximal entanglement of the emitted photon pairs ($\epsilon<1$).  
 (b) Degree of entanglement $\epsilon$ of emitted photon pairs from the elongated QDs with stress-controlled vanishing $S$ as a function of the QD elongation, $\eta$. }
\label{fig5}
\end{figure}
 
\begin{figure}[t]
\includegraphics[width=12.0cm]{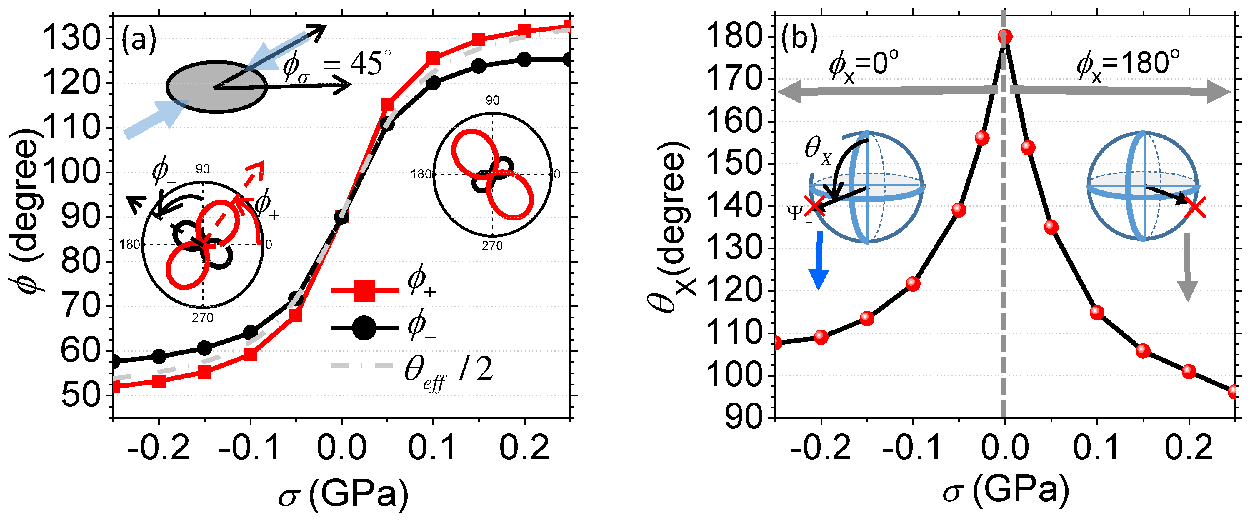}
\caption{(a) Optical polarization angles $\phi_{\pm}$, with respect to the $x$- or $y$-axes as depicted in the inset, of the excitonic fine structure states of a stressed QD with $\phi_{\sigma}=45^\circ$ as a function of the stress strength $\sigma$, which follow nearly the same $\sigma$-dependence as that of  $\theta_{eff}/2$ (See Eq.(\ref{Delta_eff_1_2}) for the definition of $\theta_{eff}$).  (b) Phase angles $\theta_{X}$ and $\phi_{X}$ used to characterize the Bloch vectors of the exciton states of the stressed QD, against the strength and orientation, $\sigma$ and $\phi_{\sigma}$, of the applied stresses, and show highly tunable by stressing the dot. }
\label{fig6}
\end{figure}

\section{Discussions}

Below, we discuss three remarkable photonic and fine structure features of stressed QDs that are revealed by the established model.
  
\subsection{Stress-dependent polarized fine structures}

Figure~\ref{fig5}(a) presents the calculated S's versus DOP's of the exciton fine structures of $x$-elongated QDs under an uniaxial stress, along the elongation axis of the QDs, with the varied stress strength from $\sigma=-0.3$ to $0.1$GPa. It is clearly seen that the $S$'s are correlated with and show quasi-linear dependences on the DOP's for a asymmetric QD with a specific elongation ($\eta=1, 1.05, 1.1$, or $1.2$). Such a $S$-DOP correlation has been noticed and inferred from the previous experiments on strained self-assembled QD systems (See Ref.\cite{Lin}). Here, with stress as an additional tunable parameter, the underlying physics in the correlated $S$'s and DOP's of  unstrained DE-QDs can be more clearly recognized. From Eqs.(\ref{optaxes1})-(\ref{intensitymax}), the degree of polarization for an exciton FS doublet of a QD under an uniaxial stress is derived as $|\text{DOP}|=|\frac{\Delta_{eff}^{xc} +\Delta_1 \cos\theta_{eff}}{E_X^S}| =|\frac{|S| + 2\Delta_1 \cos\theta_{eff}}{2 E_X^S}|$, explicitly showing the $S$-DOP relationship and explaining the linear dependences of the $S$'s on the DOP's.  
 
Furthermore, we proceed with the analysis for the effects of stress that is misaligned to the elongation axis of a QD ($\phi_{\sigma} \neq 0$). 
If the misaligned stress is so significant that the last stress-dependent term on the r.h.s. of Eq.(\ref{Deltaeff1}) is dominant ($\sigma > 0.1\text{GPa}$ in the cases studied here), one can show that the phase angle (for $0^\circ < \phi_{\sigma} < 45^\circ $),
\begin{equation} \label{thetaeffsigma}
\theta_{eff} \sim  \tan^{-1} \left( \frac{2\sqrt{3}|b|(s_{11}-s_{12})}{|d|s_{44}}\tan 2\phi_{\sigma}\right)\, . 
\end{equation} 
In other words, following Eq.(\ref{phipm}) the direction of optical polarization of a QD under a high uniaxial stress roughly follow (not exactly aligned to) the stress axes since $\phi_{\pm} \sim \frac{1}{2}  \tan^{-1} \left( \frac{2\sqrt{3}|b|(s_{11}-s_{12})}{|d|s_{44}}\tan 2\phi_{\sigma}\right) \sim \frac{1}{2}  \tan^{-1} \left( 0.8 \tan 2\phi_{\sigma}\right)$. Thus, a misaligned stress with $\phi_{\sigma}\neq 0$ leads to the optical polarization axes (with $\phi_{\pm}\neq 0$) that are misaligned to the $x-$ or $y-$axes  as well, as seen in Fig.\ref{fig2}(c). 
Equation (\ref{phipm}) and the expressions for $\delta \phi_{\pm}$ therein further predict that the two major optical axes of the exciton doublet of a QD might not be perpendicular to each other, i.e. $\delta \phi_{+} \neq \delta \phi_{-}$, which happens as $\theta_{eff} \neq 0$, e.g. as an elongated QD is subjected to a misaligned uniaxial stress to the elongation axis. Such a stress-induced non-orthogonality of the optical axes of stressed QDs has been observed in the recent study of Ref.\cite{Santosh}.

\subsection{Non-maximally entangled photon pairs from stressed QDs}

Following Eq.(\ref{Delta_eff_1_2}), the magnitude of the FSS of a QD is given by $|S|=2|\tilde{\Delta}_{eff}^{xc}| = 2\sqrt{\Delta_{eff,1}^2 + \Delta_{eff,2}^2}$ and never vanishing as long as the imaginary part, $\Delta_{eff,2}$ ($\propto \sigma\sin\phi_{\sigma}$ according to Eq.(\ref{Deltaeff2})) of the effective interaction $\tilde{\Delta}_{eff}^{xc}$ remains non-zero.  This happens as $\sigma\neq 0$ and $\phi_{\sigma}\neq 0, \pi/2$, i.e. as an uniaxial stress applied to an elongated QD is neither parallel nor perpendicular to the elongation axis.

In other words, as a prerequisite for the generation of entangled photon pair, making the FSS of a QD vanishing ($S=0$) is achievable only if the applied stress is {\it exactly} parallel or perpendicular to the axis of elongation. In the situation, the imaginary part of $\tilde{\Delta}_{eff}^{xc}$ is surely vanishing, and the resulting phase angles are $\theta_{eff}=0,\pi$. Thus, according to Eqs.(\ref{optaxes1})-(\ref{intensitymax}), one derives $\text{DOP} = \frac{2\rho_{HL}}{\sqrt{3}\Delta_{HL}}$  
and $|S|=2 |\Delta_{eff,1}|= 2|-\Delta_1 + \Delta_{VBM}|$, where $\Delta_{VBM}=\frac{2\rho_{HL} }{\sqrt{3}\Delta_{HL}}\cdot E_X^S =\text{DOP}\cdot E_X^S $.\cite{Lin}
Accordingly, a FSS is vanishing, i.e. $S=0$, only as the attractive long ranged part of {\it e-h} exchange interaction, $(-\Delta_1)$, is cancelled out by the VBM-involved repulsive interaction, $\Delta_{VBM}=\frac{2\rho_{HL} }{\sqrt{3}\Delta_{HL}}\cdot E_X^S =\text{DOP} \cdot E_X^S $.   
Re-examining Fig.~\ref{fig5}(a), one can find that the DOP's of the stress-controlled elongated QDs that are in coincidence with vanishing $S$'s are always non-zero.  
Therefore, a pair of entangled photons emitted from an elongated QDs with stress-tuned vanishing FSS should have unequal intensities and be in the so-called {\it non-maximally} entangled two-photon state, described by
$|n=2 \rangle_{ph} = (|H H\rangle +  \epsilon |V V\rangle )/\sqrt{1+\epsilon^2}$,
with the degree of entanglement,
\begin{equation}
\epsilon=\frac{ 1 - \rm{DOP}}{1 + \rm{DOP}}\, , 
\end{equation}
where $|n \rangle_{ph}$ denotes a $n$-photon state, and $H$ ($V$) indicates a $x$-($y$-) polarized photon.    Such a non-maximally entanglement ($\epsilon \neq 1$) has been shown to be advantageous for reducing the required detector efficiencies for loophole-free tests of Bell inequalities.\cite{White, Eberhard}
Figure \ref{fig5}(b) plots the degree of entanglement $\epsilon$ as a function of the elongation $\eta$ of the stressed QDs, which can be as low as $\epsilon\sim 0.6$ for $\eta=1.2$. 
 One notes that the maximal entanglement ($\epsilon=1$) is achievable only as the QD that emits the photons is perfectly symmetric so as to have DOP$=0$.\cite{Kuroda}

\begin{figure}[t]
\includegraphics[width=10.0cm]{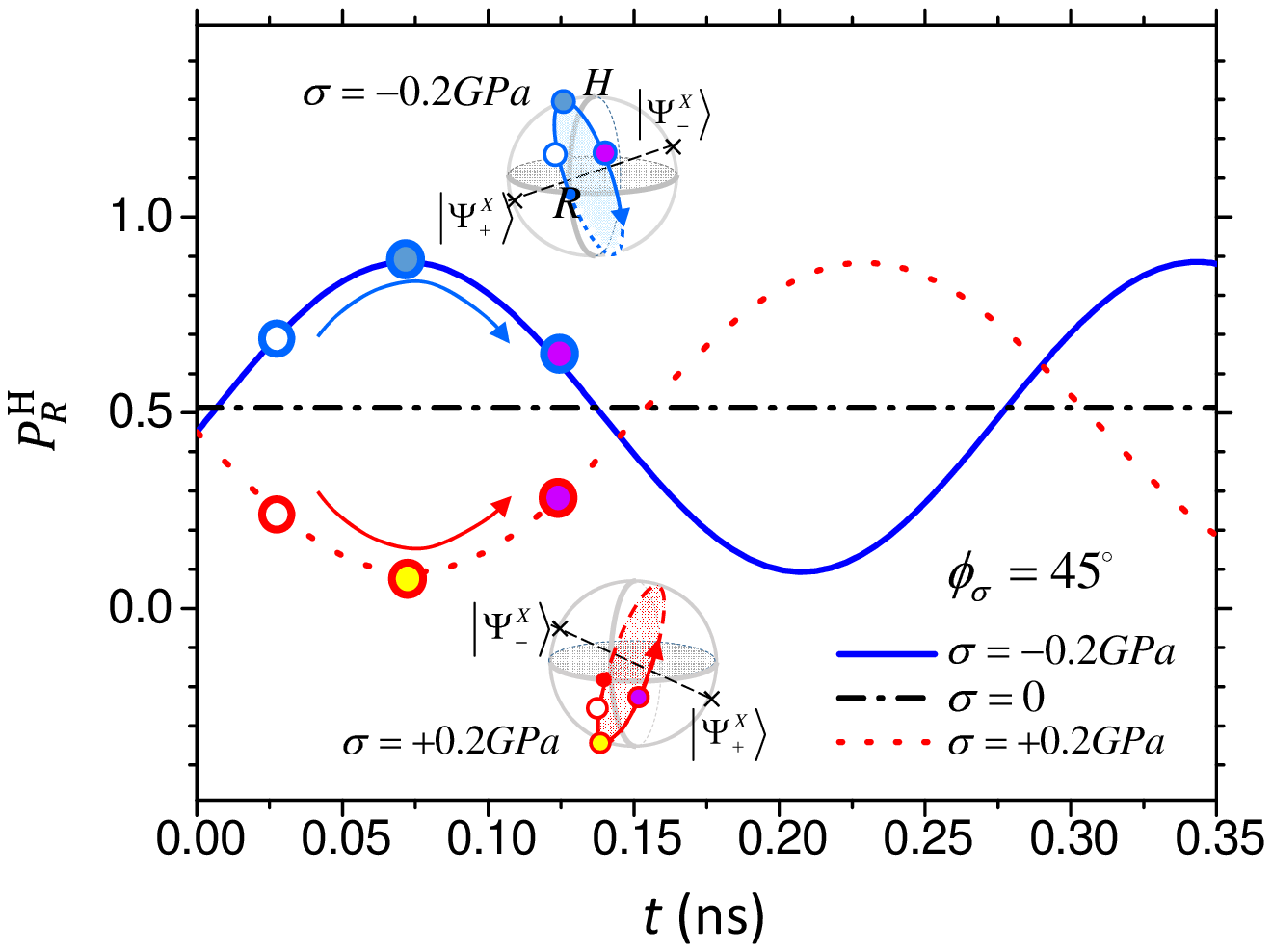}
\caption{Time dependent horizontal($H$)-polarization projection $P_R^H$ of the superposition exciton state initialized by a right-handed circularly ($R$-)polarized light for an elongated QD that is stress-free, mechanically stretched with uniaxial compressive or tensile stresses ($\sigma=\pm 0.2\text{GPa}$) along the $[100]$ direction ($\phi_{\sigma}=45^\circ$). Note that the same initialized exciton state of the QD evolves in time along distinctive paths on the Bloch spheres (See the insets) and develops very different dynamics of $P_R^H(t)$ as the applied stress is changed.}
\label{fig7}
\end{figure} 

\subsection{Mechanically prepared exciton superposition states}

Experimentally, it has been demonstrated that an exciton superposition state, $\Psi_{\hat{e}}^X=C_{+}^{\hat{e}}\Psi_{+}^X + C_{-}^{\hat{e}}\Psi_{-}^X$,in the FS of a QD can be created by a quasi-resonant laser pulse with appropriate polarization denoted by $\hat{e}$, and then evolves, within the coherence time, in a free precession, which can be geometrically represented by a circular motion on a Bloch sphere about the axis connecting the exciton eigen states.\cite{Kodriano,Poem} Uniaxially stressing a QD moves an exciton eigen state from the north or south poles of the Bloch sphere towards the equator by an angle $\theta_{X}$ with a fixed $\phi_X$ (See Eqs.(\ref{cond1}),(\ref{cond2}) and (\ref{thetaeffsigma})). Consequently, the plane of the circular motion corresponding to the free precession of the exciton state is tilted. Thus, any superposition exciton state of a QD could be optically prepared or accessed by appropriately stressing the QD prior to the optical excitation.    

Figure~\ref{fig6}(a) shows the optical polarization angles $\phi_{\pm}$ of the emission lines from the LE and HE excitonic fine structure states $\Psi_{\pm}^X$ of the QD stressed uniaxially along the $[100]$ direction ($\phi_{\sigma}=45^\circ$) as a function of the stress strength $\sigma$. One can see that, even with a fixed direction of uniaxial stress, the polarization axes rotates  over a wide angular range (almost $90^\circ$) with varying the magnitude of stress strength. Correspondingly, Fig.\ref{fig6}(b) shows the phase angles, $\theta_X$ and $\phi_X$, of the Bloch vector for the HE exciton eigen state, $\psi_{+}^X$, as formulated by Eq.(\ref{Blochv1}). One sees that the phase angles of the exciton superposition states can be related to the orientations of the optical polarizations and are roughly equal $\theta_{X}\sim 2 \phi_{+}$ ($\theta_{X}\sim 360^\circ - 2 \phi_{+}$)  for $\sigma<0$ (for $\sigma>0$), as inferred from Eqs.(\ref{cond1}), (\ref{cond2}) and (\ref{phipm}). 

The dynamics of such {\it mechanically encoded} exciton states can be monitored by optically measuring the polarization projection using the techniques presented in Refs.\cite{Benny, Kodriano}.  Figure ~\ref{fig7}(c) shows the time evolution of the $x$-polarization (or referred to as the $H$-polarization) projection $P_{R}^{H}(t)\equiv |\langle 0|P_{H}^-|\Psi_{R}^X (t)\rangle |^2 $ of the exciton superposition state that is optically initialized by a right-handed circular ($R$-) polarized laser ($|\Psi_{R}^X (t=0)\rangle = C_{+}^{R} |\Psi_{+}^{X}\rangle -i C_{-}^{R} |\Psi_{-}^X\rangle $, where $C_{\pm}^{R}=(\langle 0 |P_{R}^-|\Psi_{\pm}^X\rangle)^\ast$) of the QD uniaxially stressed in the fixed direction with $\phi_{\sigma}=45^\circ$ as considered in Fig.~\ref{fig6} (ideal coherence is assumed).\cite{Kodriano}
Analytically, one can show that $P_{R}^{H} = \frac{1}{2}[ (1+\text{DOP})^2\cos^2\phi_{+}  + (1-\text{DOP})^2\sin^2\phi_{-}  -  2 (1-\text{DOP}^2)\cos\phi_{\alpha}\sin\phi_{\beta}\sin(\frac{|S|t}{\hbar}-\phi_{+}+ \phi_{-})]$, where $\phi_{+}$ ($\phi_{-}$) is the angle of the optical axis for $|\Psi_{+}^{X}\rangle $ ($|\Psi_{-}^{X}\rangle $). 

In the absence of stress, the exciton eigen states of the $x$-elongated QD are $x$- and $y$-polarized (also referred to as $H$- and $V$-polarizations) and represented by the Bloch vectors pointing at the north and south-poles of the Bloch sphere as depicted by Fig.~\ref{fig4}(a). Thus, a $R$-polarized initial superposition state evolves on the equator around the axis connecting the north- and south-poles and the $H$-polarization projection of the temporally evolved state remains constant. Applying a compressive stress of $\sigma=-0.2$GPa to the QD tilts the plane of the free-precession circular motion by $\theta_X=110^\circ$ with $\phi_X=0^\circ$, as depicted by the schematics in the left inset of Fig.~\ref{fig6}(b) and the upper inset of Fig.~\ref{fig7}. It turns out that the $R$-polarized initial state evolves along another different path (See the inset of Fig.~\ref{fig7}), which starts from the $R$-polarized state at the equator, moves upwards but not pass the north-pole, and then turn downwards to complete the circle. Correspondingly, the $H$-polarization projection of the exciton superposition state of the compressed QD oscillates temporally at the angular frequency equal $|S|/\hbar$ as shown in Fig.~\ref{fig7}.  Similarly, a $R$-polarized initial superposition exciton state for the QD under tensile stress evolves on another circular path whose enclosed plane orientated in different direction that is titled from the north pole towards the equator by $\theta_X=100^\circ$ with $\phi_X=180^\circ$ (See the lower inset of Fig.~\ref{fig7}). As a result, the temporal oscillation of the $H$-polarization projection of the same initial exciton superposition state prepared for the same QD but with tensile stress shows to be in the opposite phase. The distinct dynamical features in Fig.~\ref{fig7} for the same QD but under different stresses indicates the possibility of the free access of any desired exciton superposition states by appropriately stressing a QD prior to an optical polarized excitation.

\section{Summary}

In summary, we present numerical investigations based on the Luttinger-Kohn four-band $k \cdot p$ theory and, accordingly, establish a valid simplified model of excitonic fine structures of droplet epitaxial GaAs/AlGaAs quantum dots under uni-axial stress control. In the formalisms, an applied uniaxial stress to a quantum dot along a specific direction acts  as a pseudo-magnetic field that is directly coupled to the pseudo-spin of exciton doublet in the fine structure of the dot, and highly tunable to tailor the level splitting and orientation of the exciton pseudo-spin.   As main results, photon pairs emitted from stressed DE-QDs are predicted always non-maximal entangled (referred to as hyper-entanglement), and a prior mechanically preparation of any desired exciton fine structure states of a QD photon source is shown feasible. The both features are associated with the valence-band-mixings in the exciton states that are especially sensitive to and controllable by external stresses for inherently unstrained droplet epitaxial quantum dots.

\section{Acknowledgements}
The authors gratefully acknowledge O. G. Schmidt, A. Rastelli, and K. Santosh (IFW Dresden) for inspiring this theoretical work. 
This work is supported by the Ministry of Science and Technology of Taiwan (Contract No. NSC-100-2112-M-009-013-MY2), and the National Center of Theoretical Sciences.

\bibliographystyle{unsrt}

\begin{thebibliography}{00}
\bibitem{parameters}
For the model calculations throughout this work, we take the parameters of QD, $l_x^e=5.9$nm, $l_y^e=5.6$nm, $l_z^e=2.9$nm, $l_x^{h}=6.6$nm, $l_y^{h}=6.1$nm, and $l_z^{h}=2.3$nm, which are extracted by fitting the numerically calculated wave functions of the QD considered in Fig.\ref{fig2}.
\end{thebibliography}

\end{document}